\documentclass[superscriptaddress,aps,pra,twocolumn,showpacs,nofootinbib,longbibliography, showkeys]{revtex4-2}
\usepackage{amsmath,amssymb,amsthm}
\usepackage{easybmat}
\usepackage[colorlinks=true,citecolor=blue,urlcolor=blue, linkcolor= magenta]{hyperref}
\usepackage[pdftex]{graphicx}
\usepackage{times, txfonts}
\usepackage{braket}
\usepackage{color}
\usepackage{natbib}
\usepackage{float}
\newcommand{\be}{\begin{equation}}
	\newcommand{\ee}{\end{equation}}
\newcommand{\ba}{\begin{eqnarray}}
	\newcommand{\ea}{\end{eqnarray}}
\newcommand{\ketbra}[2]{|#1\rangle \langle #2|}

\begin{document}
	
	%\preprint{APS/123-QED}
	\title{Impact of non-Markovian quantum Brownian motion on quantum batteries}

	\author{Gourab Bhanja\textsuperscript{}}
	\email{bhanja.1@iitj.ac.in}%Lines break automatically or can be forced with \\
	\author{Devvrat Tiwari\textsuperscript{}}
	\email{devvrat.1@iitj.ac.in }
        \author{Subhashish Banerjee\textsuperscript{}}
	\email{subhashish@iitj.ac.in }
	\affiliation{Indian Institute of Technology Jodhpur-342030, India\textsuperscript{}}

\date{\today}% It is always \today, today,
%  but any date may be explicitly specified

\begin{abstract}
Recently, there has been an upsurge of interest in quantum thermodynamic devices, notably quantum batteries. Quantum batteries serve as energy storage devices governed by the rules of quantum thermodynamics. Here, we propose a model of a quantum battery wherein the system of interest can be envisaged as a battery, and the ambient environment acts as a charger (dissipation) mechanism, modeled along the ubiquitous quantum Brownian motion. We employ quantifiers like ergotropy and its (in)-coherent manifestations, as well as instantaneous and average powers, to characterize the performance of the quantum battery. We investigate the influence of the bath's temperature and the system's coupling with the environment via momentum and position coordinates on the discharging and recharging dynamics. Moreover, we probe the memory effects of the system's dynamics and obtain a relationship between the system's non-Markovian evolution and the battery's recharging process.
\end{abstract}

%Use showkeys class option if keyword
%display desired
\maketitle

%\tableofcontents
\section{Introduction}\label{sec-intro}
The quantum theory of open systems addresses the system's dynamics, considering the ambient environment's impact~\cite{breuer2002book, weiss, sbbook}. It plays a crucial role in the dynamics of quantum mechanical systems as perfect isolation is not possible, and a complete microscopic description of the environmental degrees of freedom is not achievable or can be attainable partially~\cite{breuer2002book}. A common approach to the open system dynamics is via the master equation of the reduced density matrix of the system of interest. Time evolution in the context of the open quantum system considers the exchange of information between the system and the bath. When the system and environment time scales are well separated, this flow of information is unidirectional, that is, from the system to the environment, and is captured by the well-known GKSL (Gorini-Kossakowski-Sudarshan-Lindblad) master equation~\cite{gksl_master1, Lindblad1976}. However, the absence of this neat separation of system and environment time scales leads to a bidirectional flow of information between the system and its environment, causing the appearance of memory effects, referred to as the signature of non-Markovianity~\cite{Rivas_2014, BLP_measure, de_vega_alonso}. Recently, non-Markovian phenomena have attracted much attention in the quantum information community~\cite{Utagi2020, np_quantum_walk_Pradeep, Filippov2020, Hideaki_Matsukazi}. 

Quantum Brownian motion (QBM) is a well-known paradigm of open quantum systems, providing a unified framework wherein one can see the interconnections of some basic quantum statistical processes such as decoherence, dissipation, noise, and fluctuation. This has a rich history~\cite{bogolyubov, magalinskii, FEYNMAN1963118, CALDEIRA1983PathIA, PhysRevA.32.423, GRABERT1988115, hu_Paz_zhang, PhysRevA.69.062107, hu_matacz, PhysRevE.67.056120}. 
The Gaussian non-Markovian open system dynamics of QBM was studied in~\cite{PhysRevLett.113.200403}. This was followed up in~\cite{PhysRevLett.116.120402} and ~\cite{PhysRevA.95.020101} for two-level systems. A model of QBM bi-linearly coupled to a thermal bath~\cite{qbm_Ferialdi}, via both position and momentum couplings, which can take into account two particle pair production~\cite{huang_wei}, gives rise to non-local dissipation and rapidly relaxes towards equilibrium. Quantum to classical transition, quantum optics, Bose polaron problem, dissipative diamagnetism, stochastic thermodynamics and fluctuation theorem, single-molecule biophysics, and quantum cosmology are just a few of the many disciplines that make use of QBM~\cite{lampo2019quantum, Hanggi_2, ghosh2023quantum, Arteaga2009, hu_matacz}. Here, our objective is to characterize the generalized model of QBM from a quantum thermodynamical perspective.

Quantum thermodynamics is the study of thermodynamical processes from a quantum mechanical point of view~\cite{gemmer2004quantum, binder2019thermodynamics, sai_anders_book, deffner2019quantum, Seifert_2012, Hanggi_talkner, sekimoto2010stochastic, Alicki2018_Kosloff}. The extraction of maximal work from a system is an old problem in thermodynamics, which, in the case of quantum systems, can be quantified using the ergotropy of the system~\cite{Allahverdyan_2004, cakmak1}. Ergotropy has been established as an important quantity in the emerging field of quantum thermodynamics~\cite{coherent_ergo1, Kosloff_2013} and has recently been measured experimentally~\cite{VanHorne2020, exp_ergot2}. Due to the constant decrement in user device sizes, several thermodynamic devices, such as quantum heat engines~\cite{thomas_heat_engine, KUMAR2023128832} and quantum batteries~\cite{Binder_2015, campaioli2018quantum, PhysRevLett.118.150601, PhysRevE.87.042123}, are required to be smaller as their unit cells approach the order of molecular and atomic scales. Thus, the study of quantum batteries is fundamentally driven by the intent of perceiving devices of atomic and molecular size that could be advantageous over their macroscopic analogs~\cite{Binder_2015, campaioli2018quantum}.

A quantum battery, introduced by Alicki and Fannes~\cite{PhysRevE.87.042123}, is a system where useful energy can be stored and transferred into a thermal machine~\cite{sai_anders_book, Binder_2015, PhysRevLett.111.240401, PhysRevLett.118.150601}. The possible applications of these systems range from providing energy for operations on low-temperature quantum systems~\cite{rodriguez2022optimal} to solid-state quantum batteries~\cite{binder2019thermodynamics, Ferraro_2018}. Various theoretical bases have been implemented to review quantum batteries, including spin chains~\cite{PhysRevA.97.022106}, superconducting qubits and quantum dots~\cite{Ferraro_2018}, disordered chains~\cite{PhysRevResearch.2.023095, PhysRevResearch.4.013172} and qubits in an optical cavity~\cite{PhysRevLett.122.047702} in many body systems.

The realistic implementation of quantum batteries would need to consider dissipation and decoherence effects in the system due to unavoidable interactions with the ambient environment. Thus, studying quantum batteries in open systems has attracted much attention~\cite{Farina_2019, Kamin_2020, Carrega_2020, tiwari_2023}. Both qubits and harmonic oscillators can be fitted for the role of a quantum battery~\cite{PhysRevB.98.205423, PhysRevLett.122.047702}. To this end, we aim to study the QBM model of bi-linear coupling to a thermal bath as a model of a quantum battery, where the system of interest is the battery, and its environment provides the charger mechanism. We choose an initial non-zero ergotropy state because the environment in a thermal equilibrium state will not initially charge the quantum battery. The battery gets recharged from the environment exclusively due to non-Markovianity, and this recharging from the environment is absent in Markovian dynamics~\cite{tiwari_2023}. Further, we also aim to observe the effect of momentum coupling and bath temperature on the dynamics of the quantum battery.

The paper is detailed as follows. In Sec.~\ref{sec-non_Markovian_QBM}, we introduce the model of the QBM with generalized momentum coupling of the system with bath, and in Sec.~\ref{sec-quant_therm_character}, we discuss the characterizers of quantum thermodynamics, particularly ergotropy and their (in)-coherent parts, instantaneous and average power. This is followed by the discussion of the dynamics of the battery and the study of (in)-coherent contributions to the ergotropy in Sec.~\ref{charging_discharging}. Section~\ref{presence_of_memeory_effects} discusses the presence of memory effects in the system and its connection with the battery's dynamics, followed by conclusions in Sec.~\ref{conclusion}. 

\section{A model of non-Markovian Quantum Brownian motion}\label{sec-non_Markovian_QBM}
The QBM model to be discussed in this work was introduced in~\cite{qbm_Ferialdi}. In this model, a quantum harmonic oscillator is coupled bi-linearly to a bosonic thermal bath. The coupling of the system with the thermal bath is through a linear combination of its position and momentum operators. The total Hamiltonian $\hat H$ of the system is 
\begin{align}
     \hat{H} &= \hat{H_S} + \hat{H_E} + \hat{H_I}, \nonumber \\
     &= \frac{\hat{p}^2}{2m} + \frac{1}{2} m\omega^2_s\hat{q}^2 + \sum_n\left(\frac{\hat{p}^2_n}{2m_n} + \frac{1}{2}\omega^2_n\hat{q}^2_n\right) + (\hat{q} - \mu\hat{p})\sum_n c_n\hat{q}_n,
     \label{eq:total_ham}
\end{align}
where $\omega_s$ and $m$ are the free frequency of the harmonic oscillator and its mass, whereas $\omega_n$ is the frequency of the $n$th bath mode, respectively. Here, $\hat H_S$ can be thought of as a quantum battery and the environment $\hat H_E$ as a charger that interacts with the quantum battery via $\hat H_I$~\cite{Tabesh_Kamin2}. A similar scenario for a different model was taken up recently in~\cite{tiwari_2023}. Here \(\hat{q}\) and \(\hat{p}\) \((\hat{q}_n\) and \(\hat{p}_n\) ) are the system's ($n$th bath mode's) position and momentum operators. Furthermore, the parameter \(\mu\) provides the relative strength of the coupling with the system momentum with respect to the coupling with the system position and can be changed to the conventional position-position coupling model by canonical transformation \(\hat{x} = (\hat{q} - \mu\hat{p})\). In the case where the linear coupling of the system to the bath via its momentum operator is absent, that is, $\mu = 0$, the above model reduces to the one whose master equation was obtained by Hu, Paz, and Zhang in~\cite{hu_Paz_zhang}. Moreover, the master equation for the system coupled via position-position coupling to an Ohmic environment and in the high-temperature limit was obtained by Caldeira-Leggett~\cite{CALDEIRA1983PathIA, CALDEIRA1983Tunnel}.

We use the direct numerical method to examine the dynamics of the reduced state of the harmonic oscillator system of interest. That is, the reduced state $\rho_S(t)$ at any time $t$ ($\hbar = 1$) is given by 
\begin{equation}
    \rho_S(t) = {\rm Tr_E} \left(e^{-i\hat Ht}\rho(0)e^{i\hat Ht}\right),
    \label{systems_reduced_state}
\end{equation}
where $\rho(0) = \rho_S(0)\otimes\rho_E(0)$ is the initial joint state of the system and environment. The initial state of the environment is taken to be the thermal state $\rho_E(0) = e^{-\beta \hat H}/Z$, where $Z = {\rm Tr}\left[e^{-\beta\hat H}\right]$ is the partition function and $\beta = 1/k_B T$ is inverse of temperature. 

We envisage the harmonic oscillator system as a quantum battery discharging and subsequently recharging due to interaction with the environment. To characterize the quantum thermodynamic properties, particularly the extraction of work from the system, we make use of quantifiers such as ergotropy and power, which are briefly discussed below.  

\section{Characterizers of quantum thermodynamics}\label{sec-quant_therm_character}
Having discussed the QBM, we next undertake the task of characterizing its various quantum thermodynamic properties. This will be used subsequently to recast the QBM problem as a quantum battery. To this end, we briefly discuss the concepts of ergotropy and power.
\subsection{Ergotropy}\label{sec-ergotropy}
The maximum amount of work that can be extracted through a cyclic unitary transformation of the initial state from a quantum system is quantified by ergotropy~\cite{Allahverdyan_2004}. Considering a state governed by a time-dependent Hamiltonian $H_S + V(t)$, the transfer of work to external sources is attributed to the time-dependent potential $V(t)$. Assuming the source is connected at time $t= 0$ and disconnected at time $t = t_0$, making the process cyclic, such that $V(0) = V(t_0) = 0$. To this end, among all the final states $\rho_{t_0}$ reached from the initial state, one looks for the state with the lowest final energy. The thermal equilibrium state is a standard answer to this problem. However, in general, in the case of finite systems, only the action of $V(t)$  may not be sufficient for the initial state to reach a thermal state in time $t_0$. Therefore, in these systems, the maximum amount of work extracted (ergotropy) is smaller than the work extracted when the final state becomes a thermal equilibrium state. Below, we have outlined the method to calculate the ergotropy of a system. We assume that a quantum state $\rho_0$ and its internal Hamiltonian $H_S$ have the following spectral decomposition
\begin{equation}
    \rho_0 = \sum_{i} r_i\ket{r_i}\bra{r_i},
\end{equation}
 and 
 \begin{equation}
     H_S = \sum_{i} \epsilon_i\ket{\epsilon_i}\bra{\epsilon_i},
 \end{equation}
where ordering of the eigenvalues for $\rho$ and $H$ is in the decreasing $r_1\ge r_2\ge....,$ and increasing $\epsilon_1\le \epsilon_2\le...$ order, respectively. Due to the unitary dynamics of the state, any decrease in the system's internal energy, with respect to its self-Hamiltonian $H_S$, is extracted as work. Thus, one aims to minimize the internal energy of the final state to find the ergotropy
\begin{equation}
    \mathcal{W} (\rho_0) = {\rm Tr}(\rho_0 H_s) - \min\{{\rm Tr} (U\rho_0 U^{\dagger} H_S)\},
    \label{eq-ergotropy_1}
\end{equation}
where the minimization is performed over all possible unitaries.

The final state $\rho_f = U \rho_0 U^{\dagger}$ should commute with $H_S$ and have the same eigenvalues as the initial state. The state that achieves this minimum has the form $\rho_f = \sum_j r_j \ket{\epsilon_j}\bra{\epsilon_j}$. $\rho_f$ is the passive state, i.e., no work can be extracted from it. In this state, one can intuitively observe that the initial state's highest occupation fraction $r_1$ occupies the lowest level. The unitary operator $U = \sum_j\ket{\epsilon_j}\bra{r_j}$ performs this transformation.
Now the ergotropy of the system $\mathcal{W}(\rho_0)$ can be rewritten as 
\begin{equation}
    \mathcal{W}(\rho_0) = \sum_{j, i} r_j\epsilon_i (|\bra {r_j} \epsilon_i\rangle|^2 - \delta_{ij}).
    \label{eq-ergotropy_2}
\end{equation}
Ergotropy has been studied in the case of the open quantum systems in~\cite{cakmak1, Kamin_2020, Touil_2022, tiwari_2023}. It depends upon the initial state of the system and the system's Hamiltonian. To this end, in the case of the dynamics of the open quantum system defined by Eq. (\ref{systems_reduced_state}), we feed the state $\rho_S(t)$ as the initial state for the calculation of ergotropy in Eq. (\ref{eq-ergotropy_1}) to get the maximum work that can be extracted from the state $\rho_S(t)$ at any time $t$. 

Further, one can also account for the coherent and incoherent contributions to the ergotropy~\cite{coherent_ergo1, coherent_ergo2, tiwari_2023}. The role of quantum coherence in non-equilibrium scenarios, such as entropy production~\cite{Santos2019, Varizi_2021} and work fluctuations~\cite{Lobejko2022workfluctuations} has been discussed along with its role in thermal operations~\cite{Oppenheim_2002, Horodecki2013, Lostaglio_2015}. The incoherent ergotropy $\mathcal{W}_i (\rho)$ is the maximum work extracted from a state without changing its coherence. The expression for the incoherent ergotropy $\mathcal{W}_i (\rho)$ is given by
\begin{equation}
    \mathcal{W}_i(\rho) = {\rm Tr} \left[(\rho-\sigma)H_S\right], 
\end{equation}
where $\sigma$ is the coherence invariant state of $\rho$ with property ${\rm Tr}\left[\sigma H_S\right] = \min\limits_{\mathcal{U}\in \mathcal{U}^{(i)}} {\rm Tr}\left[\mathcal{U}\rho\mathcal{U}^\dagger H_S\right]$, where $\mathcal{U}^{(i)}$ are the unitary operators that do not change the coherence of the state $\rho$. An alternate way to calculate the incoherent ergotropy, which we have used here, is by deleting the coherence terms of the state $\rho$ and then using this dephased state to calculate the ergotropy. Therefore, incoherent ergotropy $\mathcal{W}_i (\rho)$ is given by 
\begin{equation}
    \mathcal{W}_i(\rho) = \mathcal{W}(\rho^D) = {\rm Tr}\left[\left(\rho^D - \rho_f^D\right)H_S\right],
    \label{inco_ergo}
\end{equation}
where $\rho_f^D$ is the passive state corresponding to the dephased state $\rho^D = \sum\limits_i\bra{i}\rho\ket{i}\ketbra{i}{i}$. Further, the expression for the coherent ergotropy $\mathcal{W}_c (\rho)$ is 
\begin{equation}
    \mathcal{W}_c (\rho) = \mathcal{W} (\rho) - \mathcal{W}_i (\rho).
    \label{coherent_ergo}
\end{equation}
The coherent ergotropy $\mathcal{W}_c (\rho)$ is the work that is exclusively stored in the coherence of the state. 
\subsubsection{Instantaneous and Average powers}
The instantaneous
charging power is defined by the derivative of the ergotropy as 
\begin{equation}
    \mathcal{P}(t) = \lim_{\Delta t\rightarrow 0} \frac{\mathcal{W}(t + \Delta t) - \mathcal{W}(t)}{\Delta t} = \frac{d\mathcal{W}}{dt}.
\end{equation}
The instantaneous power denotes the charging and discharging behavior of the system envisaged as a quantum battery. Positive instantaneous power indicates the charging of the battery, whereas negative instantaneous power indicates the discharging of the battery.
It is also possible to define the average power-to-energy transfer given by 
\begin{equation}
    \mathcal{P}_{av} = \frac{\mathcal{W} (t) - \mathcal{W} (t_0)}{t - t_0},
\end{equation}
where $t-t_0$ refers to the charging time of the battery. A non-zero average power indicates that the quantum battery has charged.  
\section{Dynamics of the quantum battery}\label{charging_discharging}
Now, we recast the QBM problem as a model of a quantum battery. Here, the system of interest $H_S$ (Eq.~\ref{eq:total_ham}) is the battery, and its environment $H_E$ (Eq.~\ref{eq:total_ham}) provides the charger mechanism. The environment, being in a thermal equilibrium state, will not initially charge the quantum battery. To this end, we choose an initial state of the quantum battery, which has non-zero ergotropy that dissipates to the environment. However, due to the non-Markovian nature of the environment (discussed in Sec.~\ref{presence_of_memeory_effects}), the battery gets recharged. This discharging-recharging behavior is a uniquely non-Markovian feature and will not be observed in a Markovian scenario~\cite{tiwari_2023}. 

\begin{figure}[!htb]
    \centering
    \includegraphics[width = 1\columnwidth]{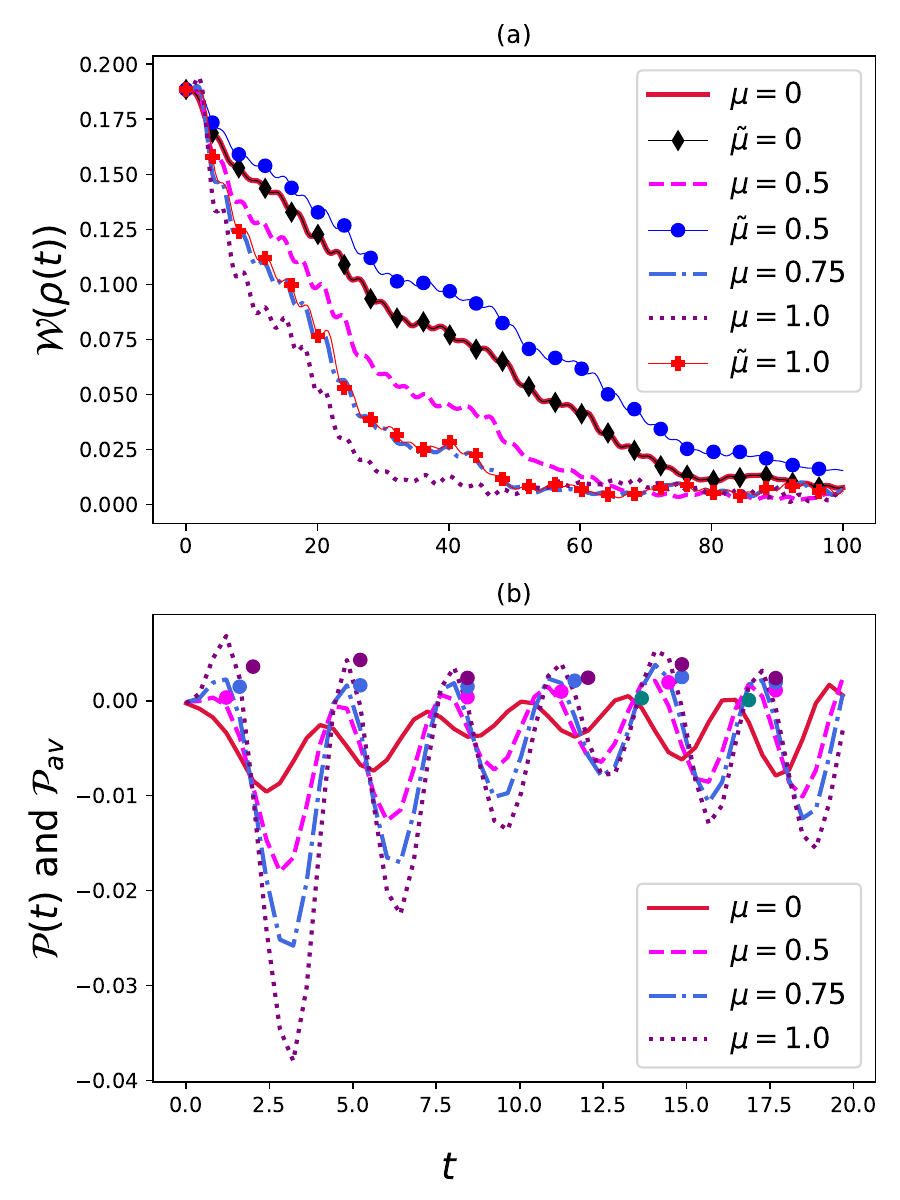}
    \caption{Variation of ergotropy $\mathcal{W}(\rho(t))$ (subplot (a)), instantaneous $\mathcal{P}(t)$ and average $\mathcal{P}_{av}$ powers (subplot (b)) with time $t$ (in natural units, where $\hbar = k_B = 1$) for different values of momentum coupling coefficient $\mu$, where we obtain $\rho_S(t)$ from Eq. (\ref{systems_reduced_state}). The dots of respective colors depict the average power during the charging cycle at a particular temperature. We have taken the following values of the parameters: $T = 1, \omega_s = 1$, and $m = 1.5$.}
    \label{fig:ergotropy_Power_vary_with_mu}
\end{figure}

We focus on the variations of the ergotropy, instantaneous, and average power of the quantum battery. The role of momentum coupling and temperature, which affects the dynamics of the quantum battery, is also studied. We take a brief pause here to emphasize the effect of the initial state $\rho_S(0)$ (in Eq. (\ref{systems_reduced_state})) of the quantum battery. It is mentioned above that we are using an initial state with non-zero ergotropy. We present here an analysis of the dynamics of ergotropy for different initial states of the battery.
\begin{figure}[!htb]
    \centering
    \includegraphics[width = 1\columnwidth]{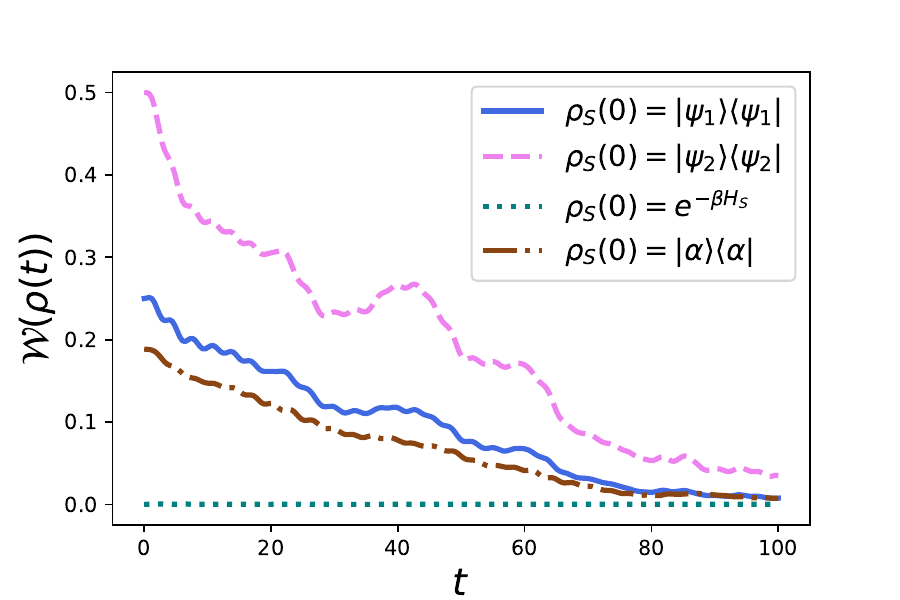}
    \caption{Variation of ergotropy $\mathcal{W}(\rho(t))$ with time $t$ (in natural units, where $\hbar = k_B = 1$) for different initial states $\rho_S(0)$ where we obtain $\rho_S(t)$ from Eq. (\ref{systems_reduced_state}). We have taken the following values of the parameters: $T = 1, \omega_s = 1$, and $m = 1.5$.}
    \label{fig:ergotropy_vary_state}
\end{figure}
In Fig.~\ref{fig:ergotropy_vary_state}, we take up different initial states of the quantum battery and plot the variation of the ergotropy. We use the following initial states for the analysis: $\ket{\psi_1} = \left(\sqrt{3}\ket{0} + \ket{1}\right)/2$, $\ket{\psi_2} = \left(\ket{0} + \ket{1}\right)/\sqrt{2}$, where $\ket{0}$ and $\ket{1}$ are the vacuum and the first excited state, respectively, for the quantum harmonic oscillator system. Furthermore, we also use the thermal equilibrium state $e^{-\beta H_S}$ and the coherent state $\ket{\alpha}$ (discussed below) for the analysis. It is evident here that the dynamics of the ergotropy depend on the quantum battery's initial state. The pattern of the variation of the ergotropy is the same for the different initial states. The ergotropy is always greater for an initial state with greater ergotropy. Further, when we take the thermal equilibrium state as the initial state of the quantum battery, we observe that the ergotropy remains zero. We choose a coherent state to analyze the thermodynamic quantities discussed in this paper due to the wide use of this state in the case of a quantum harmonic oscillator system. To this end, we take the initial state $\rho_S(0)$ in Eq. (\ref{systems_reduced_state}) to be $\rho_S(0) = \ketbra{\alpha}{\alpha}$, where $\ket{\alpha} =  e^{\alpha \hat a^\dagger - \alpha^*\hat a}\ket{0}$, with $\ket{0}$ being the vacuum state and $\alpha = 3 + 4i$. We calculate the ergotropy $\mathcal{W}(\rho(t))$ of the quantum battery for different values of the momentum coupling coefficient $\mu$ (Eq. (\ref{eq:total_ham})), which is depicted in Fig. \ref{fig:ergotropy_Power_vary_with_mu}(a).
We observe that as time increases, the ergotropy of the system decreases; that is, the amount of maximum work that can be extracted from the system decreases, and the battery discharges. Interestingly, we notice a quick drop in the ergotropy as we increase the momentum coupling coefficient $\mu$ ($\mu=0$ depicts the standard position-position coupling), indicating a rapid discharge of the battery as we raise the value of the coefficient $\mu$. 
Further, we plot the variation of the instantaneous and average powers in Fig. \ref{fig:ergotropy_Power_vary_with_mu}(b) for different values of the coefficient $\mu$. Here, we observe that the discharging and recharging rate is maximum in the case of the maximum value of $\mu$. The average power is also maximum in the case of the maximum value of $\mu$ in each charging cycle. This brings to light that the coupling of the quantum battery with the bath through its momentum along with the position coupling causes faster (dis)-charging of the quantum battery. In the absence of the momentum coupling, the model discussed here represents the exact one solved by Hu, Paz, and Zhang~\cite{hu_Paz_zhang}. In this scenario, we observe that the quantum battery discharges slowly, but it is slow while charging, too, and therefore, the average power delivered (shown using dots in Fig.~\ref{fig:ergotropy_Power_vary_with_mu}(b)) is smaller when the coefficient $\mu$ is zero. Further, the average power for zero momentum coupling appears at a much later time in comparison to the case when momentum coupling is non-zero, indicating that in the absence of momentum coupling, there is a delay in the charging of the quantum battery.

We note here that one can take a different form of interaction between the quantum battery and the bath, for example, in the place of $\hat q - \mu \hat p$ in Eq. (\ref{eq:total_ham}), one can use $(1 - \tilde \mu)\hat q + \tilde \mu \hat p$. A comprehensive investigation of the roles of the position and momentum coupling between the quantum battery and the bath can be a motivation for this. In Fig.~\ref{fig:ergotropy_Power_vary_with_mu}(a), we plot the variation of the ergotropy $\mathcal{W}(\rho(t))$ by using the latter form of interaction between the quantum battery and the bath. Here, at $\tilde \mu = 0$, there is only position-position coupling between the battery and the bath, and at $\mu = 1$, there is only momentum-position coupling between the battery and the bath. We find that the curves of the ergotropy match precisely when coefficients $\mu$ and $\tilde \mu$ are zero. Further, the values of the ergotropy are highest for $\tilde \mu = 0.5$. Interestingly, it can seen that the ergotropy is different for $\mu$ and $\tilde \mu$ equal to one. This can be attributed to the fact that in one case, the interaction involves $\hat q - \hat p$ (both position and momentum) of the battery and the bath, and in the other case, it involves only $\hat p$ (momentum) of the battery and the bath. The curve of the ergotropy for the value of the coefficient $\tilde \mu = 1$ matches with the curve of the ergotropy for $\mu = 0.75$. Thus, it points out that an overall increase in the coupling in the form of $\hat q + \hat p$ involved between the system and the bath corresponds to higher values of ergotropy. Also, the relative strength between the $\hat q$ and $\hat p$ coupling impacts the dynamics of the battery.

\begin{figure}[!htb]
    \centering
    \includegraphics[width = 1\columnwidth]{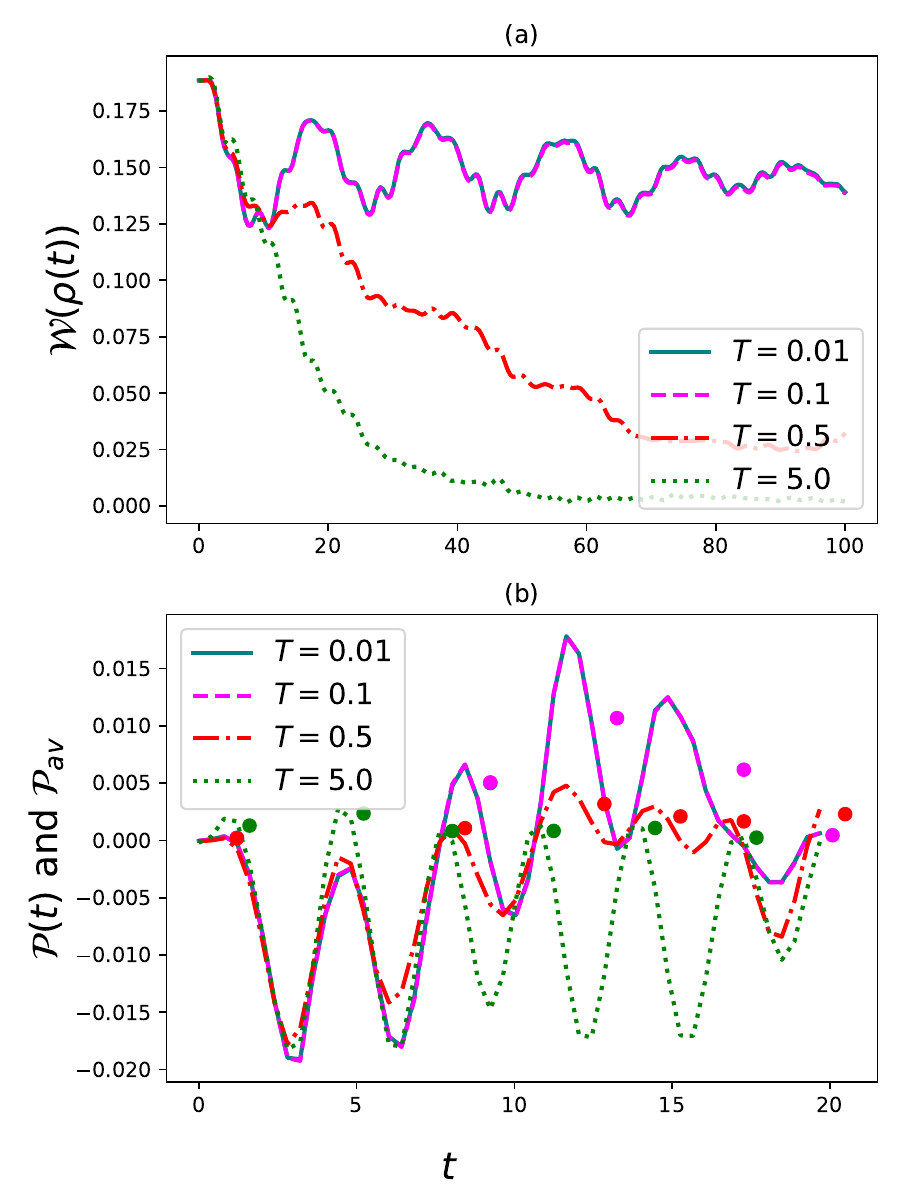}
    \caption{Variation of Ergotropy $\mathcal{W}(\rho(t))$ (in subplot (a)), instantaneous and average powers (in subplot (b)) with time $t$ (in natural units, where $\hbar = k_B = 1$) for different values of temperature $T$. The dots of respective colors depict the average power during the charging cycle at a particular temperature. We have taken the following values of the parameters: $\mu = 0.5, \omega_s = 1$, and $m = 1.5$.}
    \label{fig:ergotropy_power_vary_with_temp}
\end{figure}

We now study the effect of temperature on the dynamics of the quantum battery. To this end, we fix the value of momentum coupling $\mu = 0.5$ and use the same initial state $\rho_S(0) = \ket{\alpha}\bra{\alpha}$ (as discussed above) in Eq. (\ref{systems_reduced_state}) to obtain the dynamics of the system. This has been depicted in Fig. \ref{fig:ergotropy_power_vary_with_temp}. 
We observe that the ergotropy persists for a longer duration for lower temperatures and exhibits oscillatory behavior. The variations in the ergotropy coincide when the temperatures are 0.1 and 0.01. The ergotropy drops to zero when we increase the temperature to 5.0. Therefore, lower temperatures support the rechargeable capacity of the quantum battery from the environment, whereas higher temperatures negatively impact the performance of the quantum battery. This effect can also be observed from the variations of instantaneous and average powers plotted in Fig.~\ref{fig:ergotropy_power_vary_with_temp}(b). At lower temperatures, the instantaneous and average power achieve greater values than those in the higher temperature limit. Also, the instantaneous power amounting to the charging rate of the system is positive for a longer duration at low temperatures. 

\begin{figure}[!htb]
    \centering
    \includegraphics[width = 1\columnwidth]{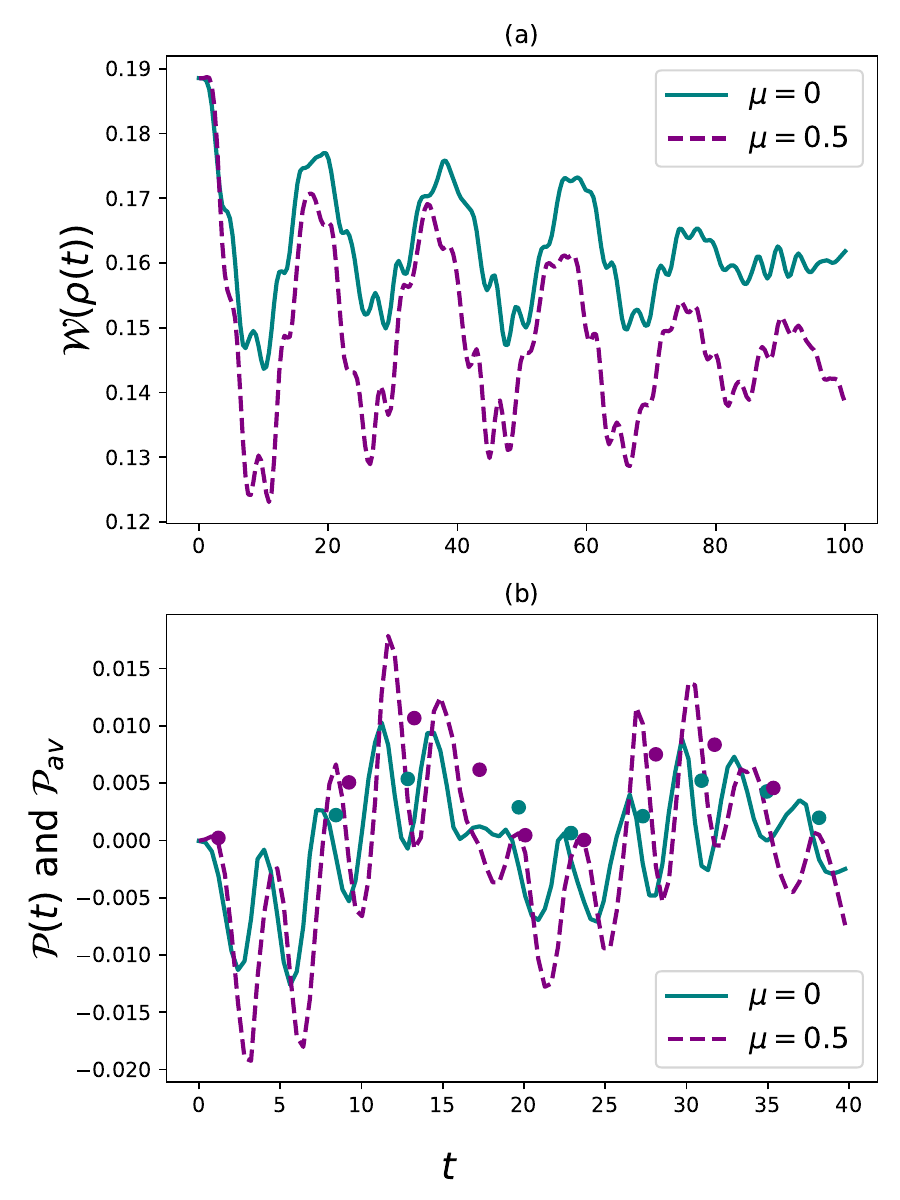}
    \caption{Variation of Ergotropy $\mathcal{W}(\rho(t))$ (in subplot (a)), instantaneous and average powers (in subplot (b)) with time $t$ (in natural units, where $\hbar = k_B = 1$) for different values of momentum coupling constant $\mu$ at low temperature $T = 0.1$. The dots of respective colors depict the average power during the charging cycle at a particular temperature. We have taken the following values of the parameters: $\omega_s = 1$, and $m = 1.5$.}
    \label{fig:ergotropy_Power_low_temp_with_mu}
\end{figure}
Furthermore, motivated by the better performance of the quantum battery at lower temperatures, we now analyze the impact of momentum coupling of the battery with the environment in the low-temperature regime. To this end, we depict the variation of the ergotropy, average, and instantaneous powers in Fig.~\ref{fig:ergotropy_Power_low_temp_with_mu} with time for a lower temperature. 
Here, we observe that the ergotropy for the momentum coupling coefficient, $\mu = 0$, is always greater than that when the battery is coupled via both position and momentum with the bath. This shows that we can extract a greater amount of work from the quantum battery when there is no momentum coupling. However, the variations in the instantaneous power for both zero and non-zero momentum coupling suggest that the rates of discharging and recharging of the quantum battery are higher for non-zero momentum coupling. Further, at various times, we observe that the average power during the charging cycle of the quantum battery is higher in the case of the non-zero momentum coupling. This analysis brings out an important facet of the system under study, that is, if we can tune the coupling of the quantum battery via its momentum with the bath, we can control the quick discharging/recharging of the quantum battery. Quick recharging followed by a slow discharge is a favorable scenario for a battery, and in this case, if we keep the momentum coupling while recharging and remove it while discharging, we can achieve a better-performing quantum battery.

We now move on to study the various components of ergotropy (namely, coherent and incoherent ergotropies) and their connection with the coherence of the quantum state. 

\subsection{The (in)-coherent ergotropy of the system and coherence of the state}
Quantum coherence is at the heart of quantum mechanics and quantum computation and has applications in practical scenarios~\cite{coherence_adesso}. Traditionally, quantum coherence depicted the presence of the off-diagonal terms in the density matrix of the state. A wide variety of measures of coherence are there, which make use of the off-diagonal terms of the density matrix. One such measure is the $l_1$ norm of coherence $\mathcal{C}_{l_1}(\rho)$~\cite{l1_norm_coherence} defined by 
\begin{equation}
    \mathcal{C}_{l_1}(\rho) = \sum\limits_{\substack{i, j \\ i \neq j}} \left|\rho_{i,j}\right|.
\end{equation}

\begin{widetext}

\begin{figure}[!htb]
    \centering
    \includegraphics[width = 1\columnwidth]{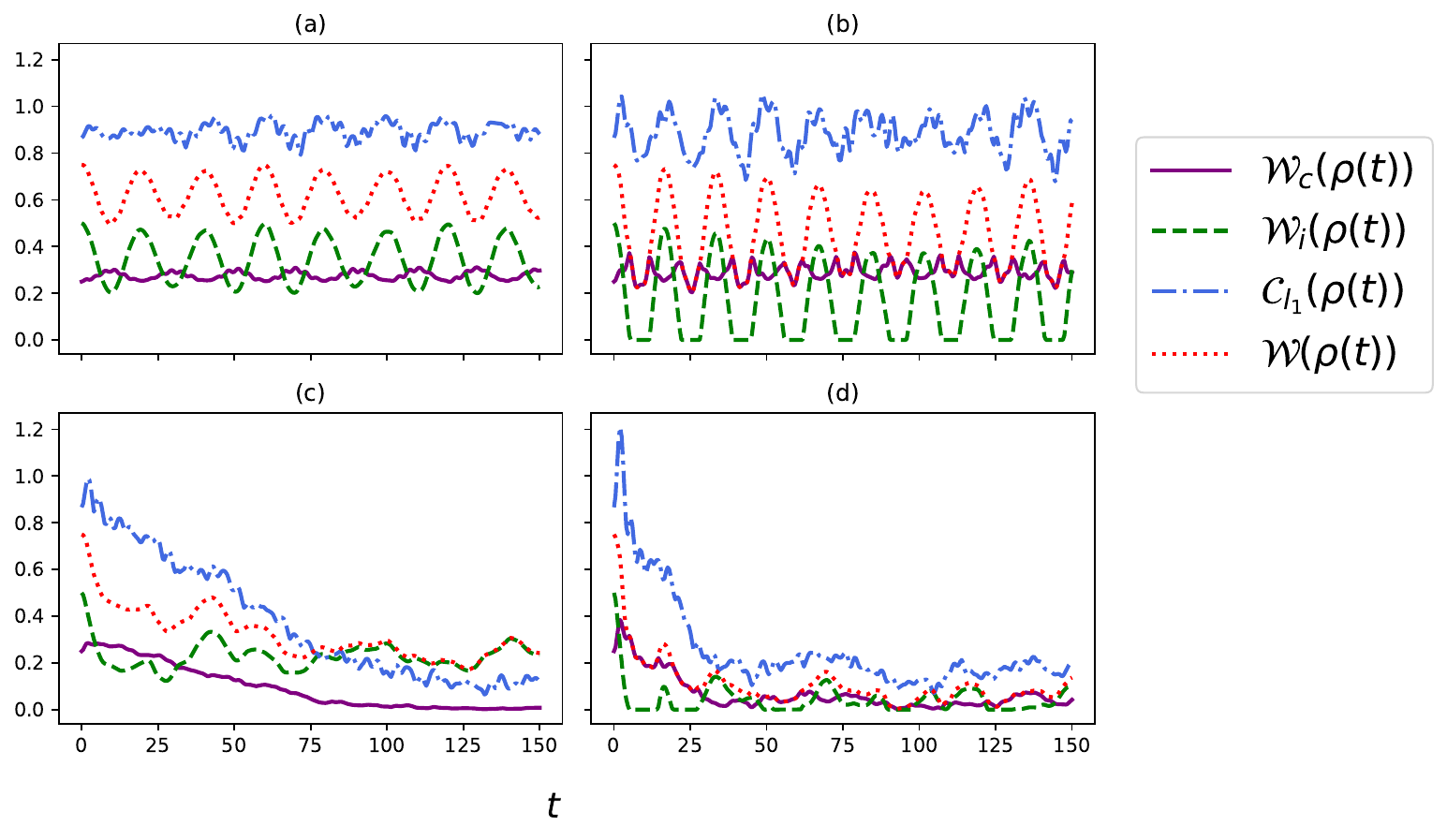}
    \caption{Variation of coherent ergotropy $\mathcal{W}_c(\rho(t))$, incoherent ergotropy $\mathcal{W}_i(\rho(t))$, $l_1$ norm of coherence $\mathcal{C}_{l_1}(\rho(t))$, and ergotropy $\mathcal{W}(\rho(t))$ with time $t$ (in natural units, where $\hbar = k_B = 1$) for different temperatures and momentum coupling coefficients $\mu$. In (a), we have $T$ = 0.1 and $\mu =0$, in (b), we have $T = 0.1$ and $\mu = 1$, in (c), we have $T = 1.0$ and $\mu = 0$, and in (d), we have $T = 1.0$ and $\mu = 1$. The parameters are: $\omega_s = 1$, and $m = 1.5$.}
    \label{fig:Coh_incoh_ergo_Cl1}
\end{figure}
\end{widetext}

Here, we calculate the $l_1$ norm of coherence along with the coherent and incoherent ergotropy of the system, defined in Eqs.~(\ref{coherent_ergo}) and~(\ref{inco_ergo}), respectively. To this effect, we take a different initial state $\rho_S(0) = \ketbra{\psi}{\psi}$ where $\ket{\psi} = \frac{1}{2}\ket{0} + \frac{\sqrt{3}}2\ket{1}$ ($\ket{0}$ and $\ket{1}$ are the ground and excited states of the system, respectively) and use Eq.~(\ref{systems_reduced_state}) to obtain the dynamics of the system. This facilitates the study of coherent and incoherent ergotropies. We plot the coherent ergotropy $\mathcal{W}_c(\rho(t))$ and the $l_1$ norm of coherence $\mathcal{C}_{l_1}(\rho(t))$ together with incoherent ergotropy $\mathcal{W}_i(\rho(t))$ and ergotropy $\mathcal{W}(\rho(t))$ in Fig. ~\ref{fig:Coh_incoh_ergo_Cl1} for different temperatures and momentum coupling coefficients $\mu$.
At lower temperatures and zero momentum coupling coefficient, we observe that the coherent ergotropy, ergotropy, incoherent ergotropy, and the $l_1$ norm of coherence all keep oscillating in similar intervals. The maxima and minima of the coherent ergotropy match with the minima and maxima of ergotropy as well as incoherent ergotropy and $l_1$ norm of coherence. However, as the momentum coupling increases, the values of the incoherent ergotropy become zero in various instances. At these times, the values of ergotropy and coherent ergotropy become equal. Further, at these points, the (in)-coherent ergotropy, ergotropy, and $l_1$ norm of coherence all attain their local minima. 

Further, at higher temperatures in Figs.~\ref{fig:Coh_incoh_ergo_Cl1}(c) and ~\ref{fig:Coh_incoh_ergo_Cl1}(d), the values of the (in)-coherent ergotropy, ergotropy and $l_1$ norm of coherence decay with time. It is interesting to note here that when the momentum coupling is zero between the system and the bath, then at longer times, the coherent ergotropy goes to zero, and in this case, the incoherent ergotropy becomes equal to the ergotropy of the system. The decay of the coherent ergotropy, as well as the $l_1$ norm of coherence, is approximately monotonic in this scenario. Meanwhile, at the same temperature when momentum coupling is higher, the incoherent ergotropy goes to zero at various points in time, followed by small revivals. At these points, the coherent ergotropy increases to equal the ergotropy of the system. 
\section{Presence of memory effects in the system}\label{presence_of_memeory_effects}
Here, we investigate the non-Markovian effects in the system defined by Eq. (\ref{systems_reduced_state}). To this end, we use the trace distance-based measure between two quantum states defined in~\cite{BLP_measure}. Consider two quantum states $\rho_1$ and $\rho_2$; the trace distance between them is
\begin{equation}
    D\left(\rho_1, \rho_2\right) = \frac{1}{2}{\rm Tr} |\rho_1 - \rho_2|,
\end{equation}
where $|A| = \sqrt{A^\dagger A}$ is the modulus of $A$. At any time $t$, the time evolution of the states is governed by Eq. (\ref{systems_reduced_state}). We consider the time evolution of the states by a family of the completely positive and trace-preserving (CPTP) maps $\Phi$, such that $\rho_i(t) = \Phi \rho_i(0)$. The trace distance gives the distinguishability of the two quantum states. 
A dynamical decrease of $D(\rho_1 (t), \rho_2 (t))$ can be interpreted as a loss of information from the open system into the environment characteristic of Markovian dynamics and vice versa. A revival of $D\left(\rho_1(t), \rho_2(t)\right)$ indicates a flow of information from the environment back to the system, signifying memory effects and non-Markovian behavior.
\begin{widetext}

\begin{figure}[!htb]
    \centering
    \includegraphics[width = 1\columnwidth]{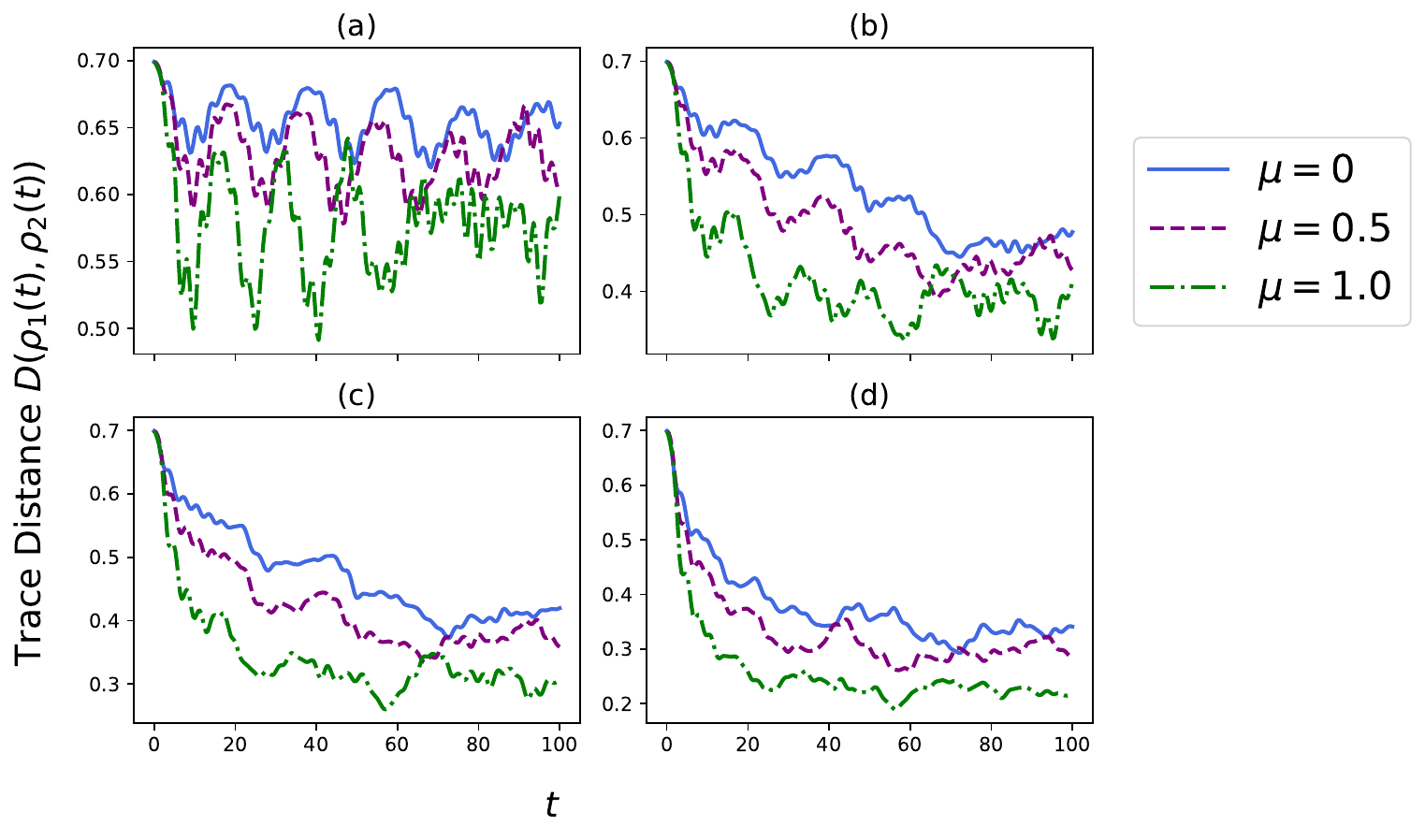}
    \caption{Variation of the trace distance $D\left(\rho_1(t), \rho_2(t)\right)$ with time $t$ (in natural units, where $\hbar = k_B = 1$) for states $\rho_1(t)$ and $\rho_2(t)$ obtained from Eq. (\ref{systems_reduced_state}) using initial states $\rho_1(0) = \ketbra{\alpha_1}{\alpha_1}$ and $\rho_2(0) = \ketbra{\alpha_2}{\alpha_2}$, where $\ket{\alpha_i} = e^{\alpha_i \hat a^\dagger - \alpha_i^*\hat a}\ket{0}$ with $\alpha_1 = 3+4i$ and $\alpha_2 = 1$, respectively. The values of temperature $T$ in (a), (b), (c), and (d) are $0.1, 0.5, 1.0$, and $5.0$, respectively. We have chosen $\omega_s = 1$, and $m = 1.5$.}
    \label{fig:trace_distance_mu_Temp}
\end{figure}
\end{widetext}
Here, we have taken two initial states $\rho_1 = \ketbra{\alpha_1}{\alpha_1}$ and $\rho_2 = \ketbra{\alpha_2}{\alpha_2}$, where $\ket{\alpha_i} = e^{\alpha_i \hat a^\dagger - \alpha_i^*\hat a}\ket{0}$ with $\alpha_1 = 3+4i$ and $\alpha_2 = 1$, respectively, and plotted the trace distance $D\left(\rho_1(t), \rho_2(t)\right)$ in Fig. \ref{fig:trace_distance_mu_Temp}. We observe that for low temperatures, the trace distance $D\left(\rho_1(t), \rho_2(t)\right)$ shows highly oscillatory behavior indicating BLP non-Markovianity~\cite{BLP_measure}, which decays as we increase the temperature. For higher temperatures, the behavior of the plots approaches monotonic behavior. In each of the subplots of Fig. \ref{fig:trace_distance_mu_Temp}, we notice that the value of the trace distance $D\left(\rho_1(t), \rho_2(t)\right)$ is lesser for higher momentum coupling coefficient $\mu$. The system shows the highest non-Markovian behavior at zero momentum coupling and low temperatures, thereby highlighting the effect of coupling of the environment to the system's position and momentum.  

Further, on comparing the variations of trace distance and ergotropy of the system, we observe that when ergotropy is higher at low temperatures and zero momentum coupling (in Figs. ~\ref{fig:ergotropy_power_vary_with_temp}(a) and ~\ref{fig:ergotropy_Power_low_temp_with_mu}(a)), the system exhibits higher non-Markovianity. However, the revivals in the trace distance $D\left(\rho_1(t), \rho_2(t)\right)$ in Fig. ~\ref{fig:trace_distance_mu_Temp} at low temperatures and non-zero momentum coupling is steep corresponding to the higher instantaneous and average powers of the system, as seen in Figs. ~\ref{fig:ergotropy_power_vary_with_temp}(b) and ~\ref{fig:ergotropy_Power_low_temp_with_mu}(b).

\section{Conclusions}\label{conclusion}
In this paper, we have studied a quantum battery modeled on the QBM coupled to a dissipative Gaussian thermal bath via position and momentum couplings. To investigate the behavior of the system of interest as a quantum battery connected to the reservoir comprised of a set of harmonic oscillators, we have studied various thermodynamical quantifiers, such as ergotropy, which is the maximum extractable work through a cyclic unitary transformation, as well as the instantaneous and average powers. Moreover, coherent and incoherent ergotropy have also been studied to account for the contribution to the ergotropy from the system state's coherent and incoherent parts.

We have investigated the effect of temperature and position-momentum coupling on the discharging-recharging behavior of the quantum battery. To this end, we observed that the ergotropy rapidly decreases with the increment in momentum coupling coefficient and temperature. It was observed that the relative strength between the $\hat q$ and $\hat p$ coupling impacts the dynamics of the battery. Further, the instantaneous power gained higher values for higher momentum coupling coefficients, indicating an improved charging rate, while lower temperatures enhanced the charging process. In the low-temperature regime, the charging can be catalyzed with the help of the momentum coupling coefficient, implying speedy charging, but it was found to impose a quick discharging, too. 
At lower temperatures, we observed the oscillatory nature of ergotropy, both coherent and incoherent ergotropy, and \(l_1\) norm of coherence, eventually decaying towards zero with time with an increase in temperature. As the momentum coupling was raised for a particular temperature, the incoherent ergotropy became zero at various instances. At these times, the values of ergotropy and coherent ergotropy became equal.

Also investigated was the presence of memory effects in the system as a benchmark for thermodynamical characteristics. In the lower temperature regime, the oscillatory behavior of trace distance provided a signature of non-Markovianity in the system and faded away monotonically with the increment of temperature as well as the momentum coupling coefficient. A qualitative agreement between non-Markovianity and thermodynamic characteristics was observed.

\section*{Acknowledgments}
S.B. acknowledges support from the Interdisciplinary Cyber-Physical Systems (ICPS) programme of the Department of Science and Technology (DST), India, Grant No.: DST/ICPS/QuST/Theme-1/2019/13.

\bibliographystyle{apsrev4-1}
\bibliography{reference}
\end{document}